\documentstyle{mn}
\input psfig.tex
\def\lsim{\lower.5ex\hbox{$\; \buildrel < \over \sim \;$}}
\def\gsim{\lower.5ex\hbox{$\; \buildrel > \over \sim \;$}}
\def \simeq{\lower.3ex\hbox{$\; \buildrel \sim \over - \;$}}
\def\ch{\lower-0.55ex\hbox{--}\kern-0.55em{\lower0.15ex\hbox{$h$}}}
\def\lh{\lower-0.55ex\hbox{--}\kern-0.55em{\lower0.15ex\hbox{$\lambda$}}}
\newif\ifAMStwofonts
\ifoldfss
  \ifCUPmtlplainloaded \else
    \NewTextAlphabet{textbfit} {cmbxti10} {}
    \NewTextAlphabet{textbfss} {cmssbx10} {}
    \NewMathAlphabet{mathbfit} {cmbxti10} {} 
    \NewMathAlphabet{mathbfss} {cmssbx10} {} 
  \fi
  \ifAMStwofonts
    \ifCUPmtlplainloaded \else
      \NewSymbolFont{upmath} {eurm10}
      \NewSymbolFont{AMSa} {msam10}
      \NewMathSymbol{\upi}     {0}{upmath}{19}
      \NewMathSymbol{\umu}     {0}{upmath}{16}
      \NewMathSymbol{\upartial}{0}{upmath}{40}
      \NewMathSymbol{\leqslant}{3}{AMSa}{36}
      \NewMathSymbol{\geqslant}{3}{AMSa}{3E}

    \fi
  \fi
\fi 
\ifnfssone
  \newmathalphabet{\mathit}
  \addtoversion{normal}{\mathit}{cmr}{m}{it}
  \addtoversion{bold}{\mathit}{cmr}{bx}{it}
  \newmathalphabet{\mathbfit} 
  \addtoversion{normal}{\mathbfit}{cmr}{bx}{it}
  \addtoversion{bold}{\mathbfit}{cmr}{bx}{it}
  \newmathalphabet{\mathbfss} 
  \addtoversion{normal}{\mathbfss}{cmss}{bx}{n}
  \addtoversion{bold}{\mathbfss}{cmss}{bx}{n}
  \ifAMStwofonts
    \ifCUPmtlplainloaded \else
      \UseAMStwoboldmath
      \makeatletter
      \new@mathgroup\upmath@group
      \define@mathgroup\mv@normal\upmath@group{eur}{m}{n}
      \define@mathgroup\mv@bold\upmath@group{eur}{b}{n}
      \edef\UPM{\hexnumber\upmath@group}
      \new@mathgroup\amsa@group
      \define@mathgroup\mv@normal\amsa@group{msa}{m}{n}
      \define@mathgroup\mv@bold\amsa@group{msa}{m}{n}
      \edef\AMSa{\hexnumber\amsa@group}
      \makeatother
      \mathchardef\upi="0\UPM19
      \mathchardef\umu="0\UPM16
      \mathchardef\upartial="0\UPM40
      \mathchardef\leqslant="3\AMSa36
      \mathchardef\geqslant="3\AMSa3E
    \fi
  \fi
\fi 

\ifnfsstwo
  \DeclareMathAlphabet{\mathbfit}{OT1}{cmr}{bx}{it}
  \SetMathAlphabet\mathbfit{bold}{OT1}{cmr}{bx}{it}
  \DeclareMathAlphabet{\mathbfss}{OT1}{cmss}{bx}{n}
  \SetMathAlphabet\mathbfss{bold}{OT1}{cmss}{bx}{n}
  \ifAMStwofonts
    \ifCUPmtlplainloaded \else
      \DeclareSymbolFont{UPM}{U}{eur}{m}{n}
      \SetSymbolFont{UPM}{bold}{U}{eur}{b}{n}
      \DeclareSymbolFont{AMSa}{U}{msa}{m}{n}
      \DeclareMathSymbol{\upi}{0}{UPM}{"19}
      \DeclareMathSymbol{\umu}{0}{UPM}{"16}
      \DeclareMathSymbol{\upartial}{0}{UPM}{"40}
      \DeclareMathSymbol{\leqslant}{3}{AMSa}{"36}
      \DeclareMathSymbol{\geqslant}{3}{AMSa}{"3E}
    \fi
  \fi
\fi 

\ifCUPmtlplainloaded \else
  \ifAMStwofonts \else 
    \def\upi{\pi}
    \def\umu{\mu}
    \def\upartial{\partial}
  \fi
\fi

\begin{document}  
  
\title{Low angular momentum accretion-outflow model of flares from Sgr A$^*$}
\author[Das \& Czerny]  
{Tapas K. Das$^1$,  Bozena Czerny$^1$\\  
 $^1$N. Copernicus Astronomical Center, Bartycka 18, 00-716 Warsaw, Poland\\ 
tapas@camk.edu.pl, bcz@camk.edu.pl
 } 
\maketitle  
\begin{abstract}  
We employ a low angular momentum accretion-outflow scenario to model the flares
emanating out from the central region of Sgr A$^*$. The primary donor for 
matter accreting onto the central SMBH of Sgr A$^*$ is assumed to be the
WR star ISR 13 E3. We analytically calculate the specific energy and angular 
momentum density of stellar wind originating from ISR 13 E3 and study the 
dynamics of that wind down to the very close vicinity of the central SMBH of 
Sgr A$^*$. We show that on the way to the Galactic centre, such wind-fed accretion may 
encounter standing shocks and such shock drives outflow from the close vicinity 
of the SMBH. Matter content of such outflow is computed and it is argued that 
such outflow is responsible for production of the Galactic centre flares. We then 
self-consistently compute the luminosity ${\cal L}_j$ and the duration time scale 
$\tau$ of such flares, as a function of fundamental accretion parameters. 
Our theoretical calculation of ${\cal L}_j$ and $\tau$ are in good agreement with
observational results.
\end{abstract}  
  
  
\section{Introduction}
\label{intro}

The center of our Galaxy harbors a massive black hole, and the surrounding
region including the central SMBH is now customarily referred as
Sgr A* as a whole, after the radio source first discovered
at that location (for a review, see e.g. Melia \& Falcke 2001).
Sgr A* shows frequent flares originating 
from the direct vicinity of the central
black hole. In the X-ray band, multiple flares are superimposed on
a steady, extended emission at the level of $\sim 2.2 \times 10^{33}$ 
erg s$^{-1}$ cm$^{-2}$ (Baganoff et al. 2003a). Two extremely bright flares 
have been reported 
so far (Baganoff et al., 2001, Porquet et al. 2003; maximum
flux of $1.0 \pm 0.1\times 10^{35} $ and $3.6^{+0.3}_{-0.4} 
\times 10^{35}$, respectively), along with 
many fainter flares
observed in the Chandra data (e.g. Eckart et al. 2004). The duration
of the flares ranges from half an hour to several hours, while 
the rise/decay
time is found to be of the order of few hundred seconds (Baganoff 2003).

Variable emission was also detected in the NIR band. Quiescence
emission takes place at the level of $\sim 1.9$ mJy (Eckart et al. 2004). 
Variable and quiescent emission
was reported by Genzel et al. (2003) based on the VLT observations, 
and by Ghez et al. (2004) based on Keck data.
In two of the events, a 17 min periodicity was found (Genzel et al. 2003).
X-ray and NIR outbursts are directly related, as shown by the detection
of a simultaneous NIR/X-ray event (Eckart et al. 2004). The 2-8 keV luminosity
of the event was $\sim 6 \times 10^{33}$ erg s$^{-1}$ cm$^{-2}$ and 2.2 $\mu$m
flux was 3.7 mJy. The duration of the X-ray event was 55 - 115 min, with 
the decay time of such flares
measured to be of the order of several minutes. 
Detection of TeV emission from the Galactic Center (Aharonian et al. 2004) 
supports the view that emission comes
from a jet.

These properties of the flare emission suggest that
such flares originate in the
innermost region of accretion flow onto the central black hole,
and we need to introduce an appropriate theoretical model of 
accretion to understand the properties of these flares. Most of the 
works on accretion processes onto the SMBH of Sgr A$^*$ concentrate 
either on almost purely Bondi (1952) type accretion, or on
 high angular momentum
ADAF type flow (Narayan,
Yi \& Mahadevan 1995; Yuan, Quataert \& Narayan 2004), 
which is sometimes claimed to 
be coupled with jet-like outflows 
(e.g. Markoff et al. 2001, Yuan, Markoff \& Falcke 2002).

In this letter, however, we employ a different kind of 
accretion model - a low angular momentum (highly sub-Keplerian) flow 
with standing shock, to explain the generation
of flares from Sgr A$^*$ and to compute the luminosity of the 
flare ${\cal L}_j$ along with its duration time scale $\tau$ in a
self-consistent way. Sub-Keplerian advective accretion onto 
galactic and extra-galactic black holes is expected to produce 
multi-transonic behaviour (see e.g. Das 2004, Barai, Wiita \& Das
2004, and references therein) and standing shocks are an essential 
ingredient in an low angular momentum flow in general
(Das 2002, hereafter D02,
Das, Pendharkar \& Mitra, 2003, hereafter DPM, and
the references therein).
Such shocks in turn play an
important role in governing the overall dynamical and
radiative processes taking place in accreting material 
and the hot, dense, exo-entropic post-shock fluid
is supposed to be responsible for launching 
jets/outflows from black hole 
accretion discs (Das \& Chakrabarti 1999, hereafter DC, 
Das, Rao \& Vadawale 2003, hereafter DRV, and references 
therein). Such a coupled accretion-outflow model may 
reproduce the characteristic behaviours of the Sgr A$^*$
flares in a natural way.
\begin{figure}
\vbox{
\vskip -0.5cm
\centerline{
\psfig{file=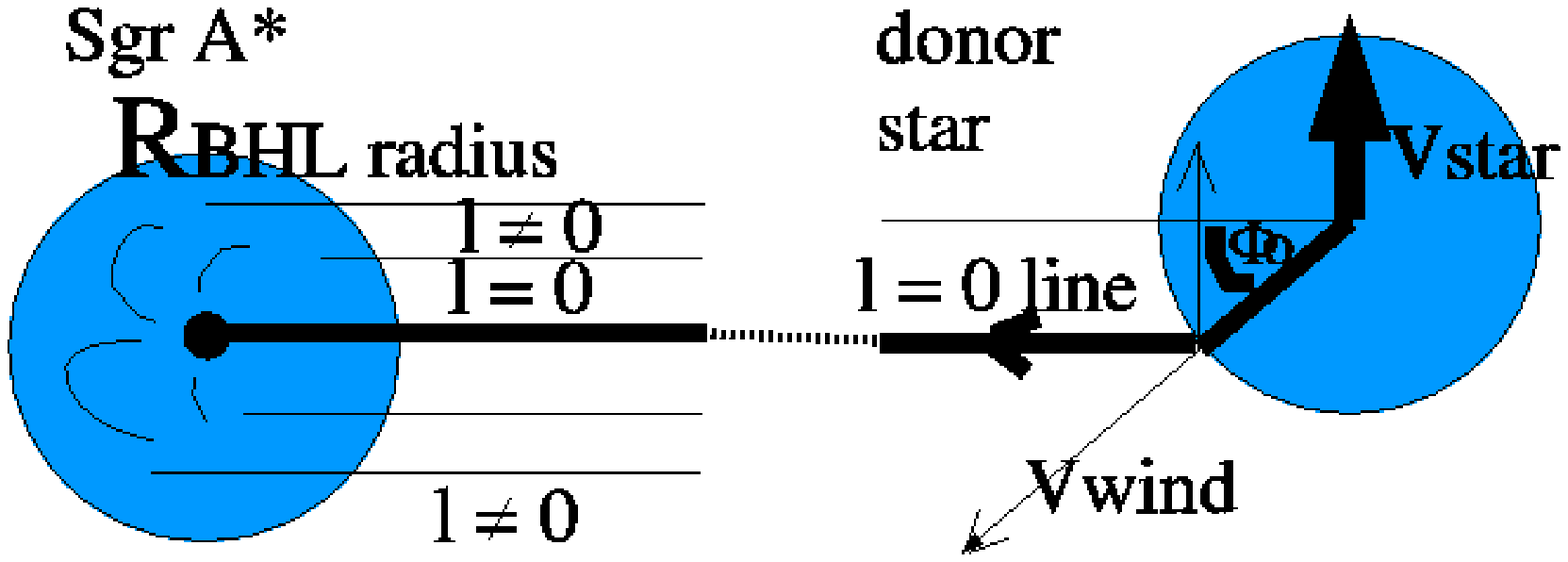,height=4cm,width=10cm,angle=0.0}}}
\noindent {{\bf Fig. 1:}
The schematic picture of the wind flow from the star IRS 13E3
to-wards the Galactic center.
}
\end{figure}
\section{Source of the accreting material}
Stellar winds originating from the central cluster are plentiful sources of 
gas. Most of this material is likely to be expelled from the central region
(e.g. Quataert 2003) but a remaining fraction may power the observed activity.
Among various mass loosing stars, particularly active one is the Wolf-Rayet star
IRS 13 E3 (Paumard et al. 2001, Melia \& Falcke 2001). We assume this star to be 
the
dominant source of the matter accreting onto Sgr A$^*$.
We determine the angular momentum and the Bernoulli constant of the flow
from IRS 13 E3 in the following way:

We consider the case where the 
wind velocity is much higher than the orbital velocity 
of a star. 
Therefore, the material
ejected from a fractional region of the star surface located at $\phi_o$ 
(see Fig.1) 
can reach the gravity center with zero angular momentum (see e.g. Loeb 2004). 
This angle is given by the condition:
\begin{equation}
\sin \phi_o = {v_{\rm star} \over v_{\rm wind}}.
\end{equation}

The wind is not significantly accelerated as long as the gravity field is dominated 
by the stellar component. It is moderately supersonic, with the Bondi-Hoyle-Lyttleton
accretion radius, $R_{\rm BHL}$, given by the formula
\begin{equation}
R_{\rm BHL} = {2 G M \over v_{\rm wind}^2 + v_{\rm s}^2},
\end{equation}
where $v_{\rm wind}$ is the flow velocity and $v_s$ is the sound speed.
We further assume that the velocity $v_{\rm wind}$ is unmodified down 
to a distance $R_{\rm BHL}$. The value of the Bernoulli constant, $\cal E$,
\begin{equation}
{\cal E} = 0.5 v_{\rm wind}^2 + { v_s^2\over \gamma_{i}-1} - 
{GM \over R_{\rm BHL}-r_{\rm g}}
\label{eq:bernoulli}
\end{equation}
is effectively determined by the value of the polytropic index on the inflow, 
$\gamma_{i}$, and an assumed gas temperature, where
$r_g=2GM_{\rm BH}/c^2$. Through out this work,
we use the Paczy\'nski and Wiita (1980) potential 
to describe the flow.

The spherically-symmetric wind blowing at $\phi \ne \phi_o$ or out of the orbital 
plane will posses certain 
amount of positive or negative specific angular momentum
(angular momentum density), $\lambda $. In the second order
approximation 
\begin{equation}
\lambda \approx -v_{\rm wind} D \delta \phi \cos(\phi_o) [1 - 1/2 \tan \phi_o \delta \phi],
\label{eq:l_loc}
\end{equation}  
where $\delta \phi = \phi - \phi_o$, $\phi$ is the azimuthal angle of the element at the
star surface and $D$ is the distance between the star and the
Galactic center. 

A cylindrical fraction of this flow, with $\Delta \phi = R_{\rm BHL}/D$, and $\Delta \theta
\sim R_{\rm BHL}/D$ will be intercepted by the central black hole, 
where the angle $\theta$ determines the deviation from the orbital plane. 

Integrating the Eq.~\ref{eq:l_loc} with respect to 
$\delta \phi $ and $\delta \theta$ in
the limits specified by $\Delta \phi$ and $\Delta \theta$,
 we obtain the net angular
momentum of the flow as:
\begin{equation}
\lambda_{\rm eff} = {3 \over 2 \pi} {1 \over \sqrt{1 - (v_{\rm star}/v_{\rm wind})^2}} v_{\rm star}D 
\left({R_{\rm BHL} \over D}\right)^2,
\label{eq:leffzero}
\end{equation}
The above relation is 
valid if the star velocity is significantly smaller than the wind velocity. Large value of
the angular momentum density of the donor star, $v_{\rm star}D$, is decreased by small 
quadratic term in the $R_{\rm BHL}/D$ ratio.

The formula changes if the wind is not perfectly isotropic. If the departure of 
the mass flux, $\delta (\rho_{\rm wind}v_{\rm wind})/<\rho_{\rm wind}v_{\rm wind}>$, happen at distances
$\delta \phi R_{\rm star}$, the net angular momentum density can be estimated by integrating
the formula ~\ref{eq:l_loc} as
\begin{equation}
\lambda_{\rm eff} = {1 \over \pi} D v_{\rm star}\left({R_{\rm BHL} \over D}\right){\delta (\rho_{\rm wind}v_{\rm wind}) \over <\rho_{\rm wind}v_{\rm wind}>},  
\end{equation}
i.e. the net angular momentum density is a fraction of $R_{\rm BHL} v_{\rm wind}$.  
The flow, initially one-sided, becomes roughly spherical below  $R_{\rm BHL}$. 
We further express $\lambda$ in dimensionless units $2 GM/c$, 
and the radius in $r_{\rm g}$.

\section{Low angular momentum flow close to Sgr A*}
If the accreting material
is assumed to be at rest far from black hole, the flow must exhibit
transonic behaviour in order to satisfy the inner boundary 
conditions imposed by the
event horizon. Low
angular momentum flow may posses
more than one sonic point, as first shown
by Abramowicz \& Zurek (1981). Typically the external
sonic point, $r_{\rm out}$, lies close to the 
corresponding Bondi radius.
The internal sonic point $r_{\rm in}$ and the middle 
sonic point $r_{\rm mid}$ exist within and outside the marginally 
stable orbit, respectively, for general relativistic (Das 2004,
Barai, Das \& Wiita 2004) as well as for post-Newtonian 
(D02, DPM) model of accretion flow.
The location of the sonic points can be calculated as a function
of the specific flow energy ${\cal E}$ (the Bernoulli's constant),
angular momentum $\lambda$ and inflow polytropic index $\gamma_i$
(see, e.g. \S 3 of D02 for details of such calculations).

If $\lambda$ is almost zero, a shock does not form, and 
accretion remains supersonic
down to the event horizon after it crosses $r_{\rm out}$.
For slightly larger $\lambda$ the centrifugal barrier 
becomes strong enough, inflowing matter starts pilling up 
close to the black hole due to the resistance offered by 
the barrier, and the depleted matter may break the incoming 
flow behind it and consequently a shock forms. Such shocks 
may become steady and standing so that they can be studied
within the framework of stationary flow (D02, DPM and references
therein).

Following D02, we consider here a stationary, non-self-gravitating, 
non-magnetized, inviscid accretion of polytropic
fluid. We assume that the flow proceeds through a 
standing shock, so the discontinuity in the radial velocity
allows to match the supersonic flow through $r_{\rm out}$ with the
subsonic flow through $r_{\rm in}$. 
The exact location of the shock as a function of
parameters $\left[{\cal E},\lambda,\gamma_i\right]$, is obtained by solving 
the generalized Rankine-Hugoniot conditions. We assume that the shock
is non-radiating and infinitesimally thin.

Formation of the shock
leads additionally to the generation of an outflow, and the shock location, $r_{\rm sh}$,
 may be assumed to be the radial length scale of
the outflow launching zone. 
The exact amount of the outflowing material ${\dot M}_{\rm out}$ can be calculated
as a function of $\left[{\cal E},\lambda,\gamma_i,\gamma_o\right]$, where
$\gamma_o$ is the outflow polytropic index and is always less that $\gamma_i$
due to the radiation momentum deposition on slowly expanding outflowing matter 
at shock location.
In our work, we assume
$\gamma_o$ to be a free parameter (subjected to the
constraint $\gamma_o < \gamma_i$), although in reality
$\gamma_o$ may be directly related to the
heating and cooling processes taking place in post-shock
matter; see DC and DRV for details.

Flares observed from Sgr A* can be due to the emission from this outflowing 
material.
Jet emission seems to be the plausible origin of both the X-ray and the NIR 
radiation (e.g. Markoff et al. 2001), if the dissipation within the shock
itself does not lead to a strong emission. In this case the luminosity of
Sgr A* during the flare, ${\cal L}_j$, is related to the outflow rate, 
${\dot M}_{\rm out}$, as ${\cal L}_j = 0.1 c^2 {\dot M}_{\rm out}$. Here we assume the
10\% efficiency of energy conversion, since studies of AGN jet indicate rather
high efficiency of jets even if the radiative efficiency of the accretion flow 
is low (e.g. Maraschi \& Tavecchio 2003).
We basically model a stationary situation. However, the flow pattern with an 
outflow can fully develop under the condition that the accretion phase with
a shock lasts long enough. Therefore, the minimum duration of the flare,
$\tau$, is associated with the infall time
scale of post-shock accretion and can be defined as:
\begin{equation}
\tau = \int_{r_{\rm sh}}^{r_e} \frac{dr}{u(r)}
\end{equation}
where $u(r)$ is the dynamical flow velocity and $r_e$ is the location of the 
event horizon. 

It is to be noted that multi-transonic flow and standing shocks form for 
a specific region of parameter space spanned by $\left[{\cal E},\lambda,
\gamma_i\right]$ (see, e.g. Figure 4. of D02 for a global classification of
shock forming parameter space). For certain values of $\left[{\cal E},\lambda,
\gamma_i\right]$, no multi-transonic or standard mono-transonic
stationary solution exists.  The corresponding flow pattern becomes inherently 
time-dependent. Such non-stationary flow might even better account for
Sgr A* variability but we have no adequate description of such a flow.
 
\begin{figure}
\vbox{
\vskip -4.5cm
\centerline{
\psfig{file=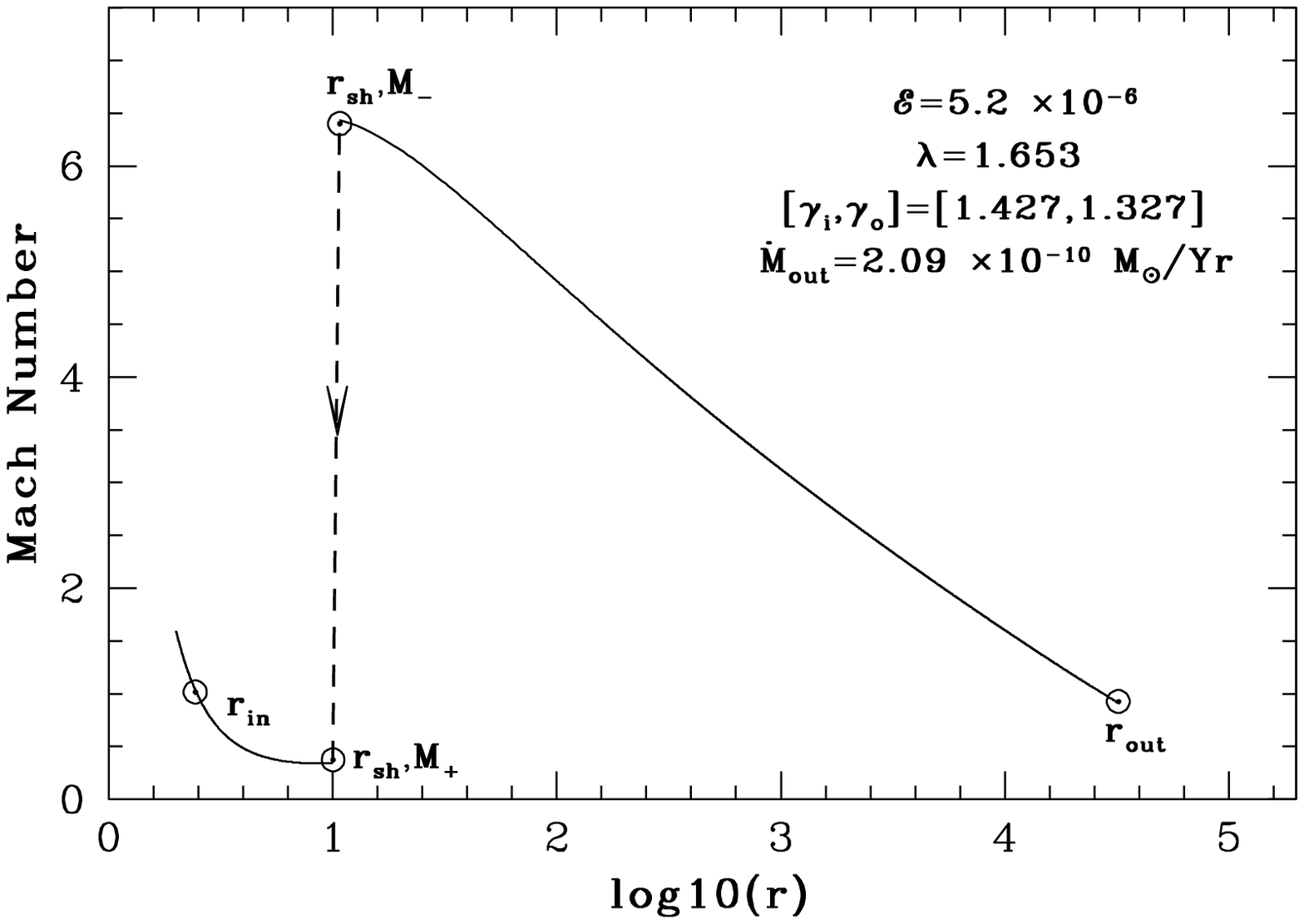,height=12cm,width=12cm,angle=0.0}}}
\noindent {{\bf Fig. 2:}
Example of a typical flow topology with a shock. The position of the shock, $r_{\rm sh}$,
is determined by the Rankine-Hugoniot conditions.
}
\end{figure}

\section{Results}

We adopt $3.8 \times 10^6 M_{\odot}$ for the black hole mass in Sgr A* 
(Ghez et. al. 2003) and
consider the accretion rate to be equal to 10$^{-8} M_{\odot}$ yr$^{-1}$. 
The distance
to the donor star is 3.5 pc,
and the wind velocity of the star, $v_{\rm wind}$, is of order of 1000 km/s (adopted after 
Rockefeller et al. 2004).
We assume that the orbital star velocity, $v_{\rm star}$, is 200 km/s, of order of 
radial velocities 
of other stars at similar distance from the center 
although the actual {\it radial} velocity 
of this star is lower (Eckart \& Genzel 1997, Paumard et al. 2001). 
\begin{figure}
\vbox{
\vskip -4.8cm
\centerline{
\psfig{file=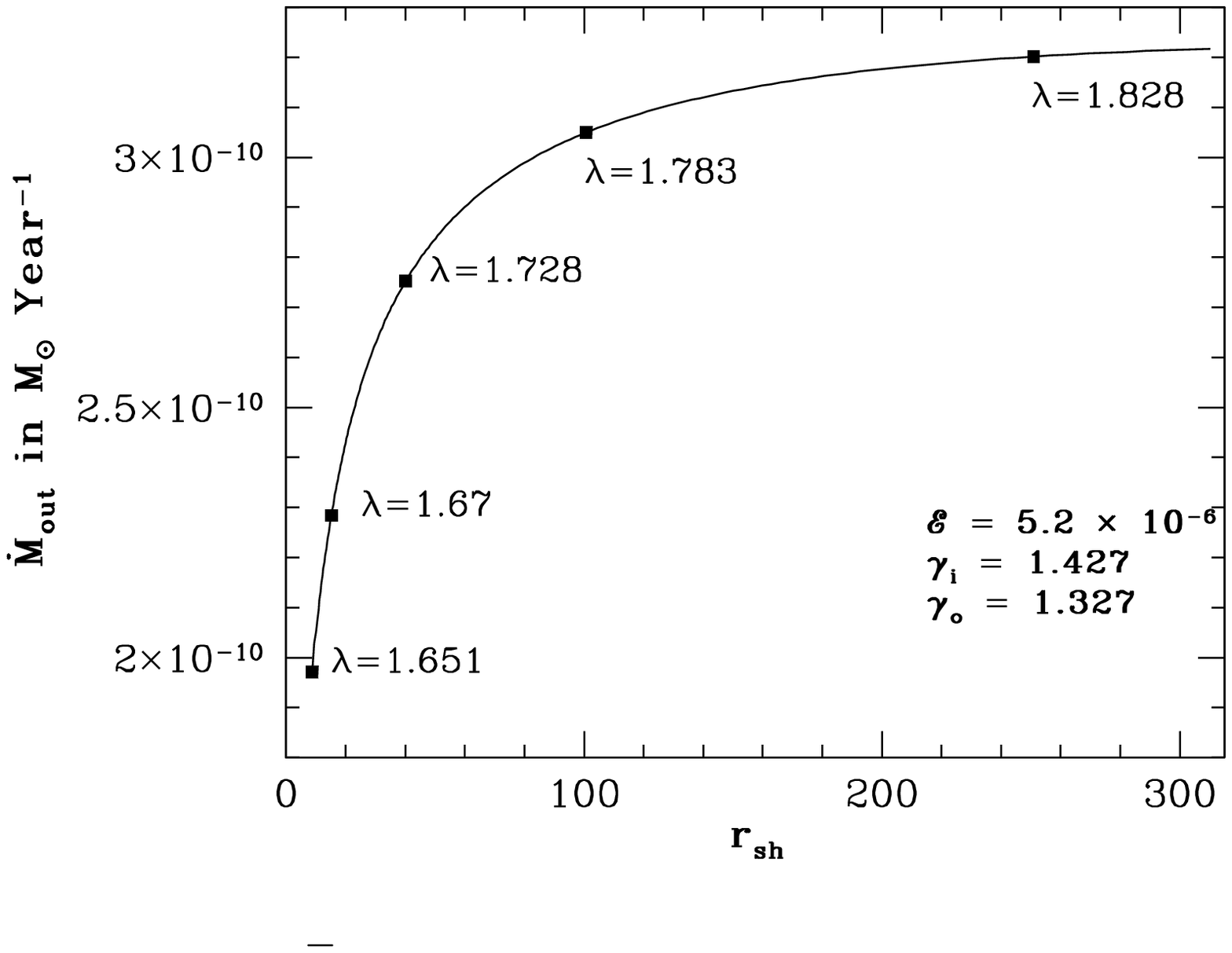,height=12cm,width=12cm,angle=0.0}}}
\noindent {{\bf Fig. 3:}
The relation between the shock location and the outflow rate. Marked along 
the curve are the corresponding values of the angular momentum density in $2GM/c$ units.
}  
\end{figure}  
 As the polytropic index of the inflow, we take the value $\gamma_{\rm i} = 1.426$,
representative for non-relativistic flow, and for the wind temperature we take 1.3 keV,
the plasma temperature estimated at the basis of the analysis of the extended X-ray 
emission (Baganoff et al. 2003a).
This means that the flow is moderately supersonic at the $R_{\rm BHL}$ ($v_s = 600$ 
km s$^{-1}$, $R_{\rm BHL} = 7.4 \times 10^{16}$ cm or $6.6 \times 10^4 r_{\rm g}$).
These values allow us to estimate the Bernoulli constant of the flow, 
${\cal E}$, in Eq.~\ref{eq:bernoulli} as
$5.2 \times 10^{-6}$ in units of $c^2$.  
The effective angular momentum given by Eq.~\ref{eq:leffzero} 
is usually small, $\lambda_{\rm eff} = 
0.14$. This means that typically the flow can proceed directly toward the black hole,
without the need for angular momentum loss and without considerable dissipation.
However, if the wind is occasionally non-uniform, 
$\delta (\rho_{\rm wind}v_{\rm wind})/<\rho_{\rm wind}v_{\rm wind}>$ of order of 2 per cent,
 changes
the net angular momentum by a factor of 11. Such a large value of the angular momentum
forms effective centrifugal barrier and the inflow may proceed through a shock, with
accompanying outflow and a burst of radiation in the X-ray and NIR bands. 

Fig. 2 represents a characteristic topology of shocked accretion flow. The 
values of $\left[{\cal E},\lambda,\gamma_i,\gamma_o\right]$ used are provided
in the figure. Matter first passes through $r_{\rm out}$ ( $3.2 \times 10^4 r_g$) and
encounters a shock at $r_{\rm sh}$ (10.03 $r_g$) close to the black hole.
The dashed vertical line marked with a down-ward arrow represents the 
shock transition.
$M_-$ and $M_+$ are the pre/post shock Mach numbers and the shock strength 
${\cal S}$ is defined as ${\cal S} = \frac {M_-}{M_+}$, which comes out to be
16.64 for this case. Post-shock subsonic
inflow becomes supersonic again after crossing $r_{\rm in}$ (2.61 $r_g$) and
finally dives through the event horizon $r_e$. Part of the shock compressed
hot and dense matter with polytropic index $\gamma_o$ (=1.327)
emerges as outflow. 
The outflow rate ${\dot M}_{\rm out}$ in this case comes out to 
be $2.1\times10^{-10} M_{\odot}$ yr$^{-1}$. The corresponding values of
${\cal L}_j$ and $\tau$ comes out to be 
1.1$\times 10^{36}$ erg s$^{-1}$ cm$^{-2}$ and $4 \times 10^3$ s respectively.

The position of the shock 
is quite 
sensitive to the choice of $\lambda$,
if other parameters of the flow are kept fixed. 
One can have a range of $r_{\rm sh}$ and  ${\dot M}_{\rm out}$ by varying 
$\left[{\cal E},\lambda,\gamma_i,\gamma_o\right]$ as well as by varying $\lambda$ 
{\it only} by keeping $\left[{\cal E},\gamma_i,\gamma_o\right]$ constant. 
For a fixed set of $\left[{\cal E},\gamma_i,\gamma_o\right]$ shown in the 
figure, we represent the dependence of the outflow rate on the location of 
the outflow launching zone in Fig. 3 by varying the value of flow angular 
momentum. Similar figure can be drawn for other appropriate values of 
$\left[{\cal E},\gamma_i,\gamma_o\right]$ as well. Note, however, that 
although very large values for $r_{\rm sh}$ (outflow launching zone) can be 
obtained as a consistent mathematical solution, they may not correspond to
the real physical situation. 
Smaller values of the outflow launching zone length scale could be obtained by
further decreasing the value of $\gamma_i$. 

Fig. 4 shows the relation between the luminosity of the 
flare ${\cal L}_j$ (in CGS unit) and the flare duration time scale $\tau$ 
(in units of one thousand seconds). The figure is drawn for the same 
$\left[{\cal E},\gamma_i,\gamma_o\right]$ used to draw the Fig. 3. 
We show the values of ${\cal L}_j$ and $\tau$ for the range of $r_{\rm sh}$ as
$r_{\rm sh}$ = 8.69 $r_g$ to $r_{\rm sh}$ = 17.03 $r_g$, the corresponding range 
for $\lambda$ is from 1.651 to 1.675. The
low luminosity flares are of shorter duration and the length scale
at which they are generated is also shorter, this is because the dynamical time 
is shorter as matter gets closer to the Black hole. 
One can have a further small (compared to the lowest value shown in the figure)
value of $\tau$ by fine tuning the values of 
$\left[\lambda,\gamma_i,\gamma_o\right]$ for a fixed value of
${\cal E}$. 
Note, however, that $\tau$ co-relates with the black hole mass. If one considers
$M_{BH} = 2.6\times 10^6 M_{\odot}$ (the lower limit of $M_{BH}$ for 
Sgr A$^*$), all values of $\tau$ will be reduced. If $\tau_{2.6}$ and
$\tau_{3.8}$ corresponds to $M_{BH} = 2.6\times 10^6 M_{\odot}$ and
$M_{BH} = 3.8\times 10^6 M_{\odot}$ respectively, then 
$\left(\tau_{3.8}-\tau_{2.6}\right)$ comes out to be about 
1250 seconds for low luminosity flares and the difference 
increases for flares with higher value of luminosity, i.e., for larger 
value of $r_{\rm sh}$.
For example, for $r_{\rm sh} = 17.03 r_g$, $\tau_{2.6}$ is only 
equal to $\sim$ 7449 seconds whereas $\tau_{3.8}$ comes out to be 
equal to $\sim$ 10887 seconds.
\begin{figure}
\vbox{
\vskip -4.5cm
\centerline{
\psfig{file=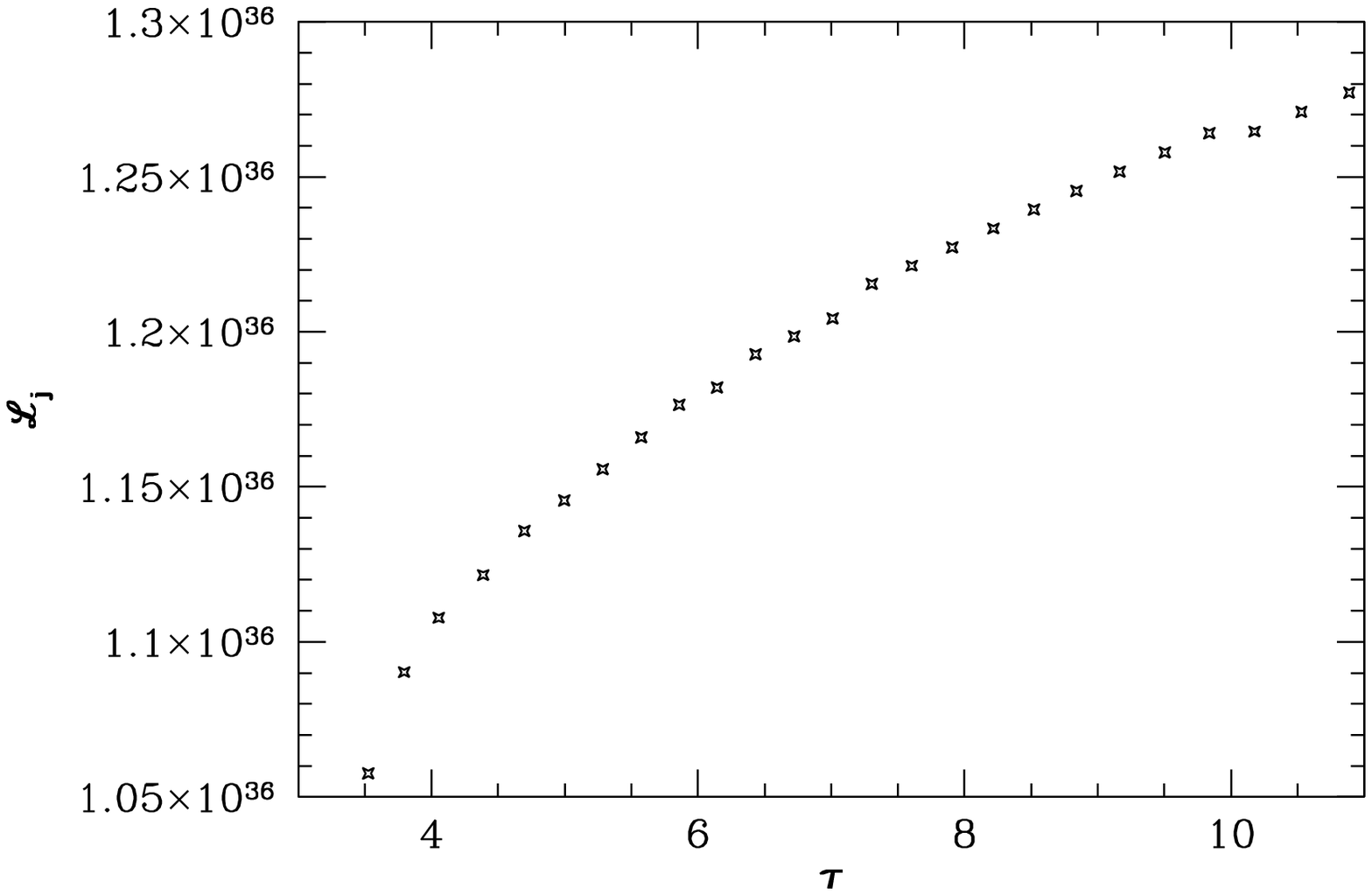,height=10cm,width=10cm,angle=0.0}}}
\noindent {{\bf Fig. 4:}
The relation between the flare luminosity ${\cal L}_j$ and
flare duration time scale $\tau$ (in thousands of seconds) for some representative cases,
see text for detail.
}
\end{figure}

\section{Discussion and conclusions}

The type 
of weakly rotating flows presented in this letter,
 have not been theoretically explored much in the literature; although they are well exhibited in
nature for various real physical situations like detached binary systems
fed by accretion from OB stellar winds (Illarionov \&
Sunyaev 1975; Liang \& Nolan 1984), semi-detached low-mass
non-magnetic binaries (Bisikalo et al.\ 1998) and super-massive BHs fed
by accretion from slowly rotating central stellar clusters
(Illarionov 1988; Ho 1999 and references therein). Even for a standard Keplerian
accretion disc, turbulence may produce such low angular momentum flow (e.g. Igumenshchev
\& Abramowicz 1999, and references therein).

We consider this type of flow as an attractive model of Sgr A* activity. We 
expect that accretion onto the central black hole proceeds roughly continuously
but small (up to 2-3 \%) fluctuations in the wind density or velocity lead to small variations 
in the angular momentum density which in turn results in temporary shock/jet
formation  and consequently an enhanced dissipation. 
The range of predicted shock positions $ 10 r_{\rm g}<r_{\rm sh} < 300 r_{\rm g}$ give
interesting range of expected burst timescales, and more luminous flares are expected to
last rather longer than fainter flares.

The predicted exemplary flare luminosities are in the interesting range.
The exact estimate of the flare bolometric luminosity is rather difficult,
as nicely discussed by Aharonian \& Neronov (2004).
As on order of magnitude estimate we can assume that the bolometric
luminosity
of the flare is ten times higher than the measured X-ray flux so the
predicted outburst would have the measured X-ray luminosity around
$10^{35}$ erg s$^{-1}$ cm$^{-2}$. 

The shown luminosity range is, however, rather narrow.
The barionic load of the outflow (and hence, ${\cal L}_j$)
is controlled mainly
by the post shock thermal pressure
and post shock thermal energy generated is modulated by the
total energy content of the flow as well as the thermal
energy content of fluid.
Hence broader luminosity range can be obtained
if we allow for variation of $\gamma_{\rm i}, \gamma_{\rm o}$ and
${\cal E}$ on the top of perturbations of the angular momentum
density. We found that ${\cal L}_j$ correlates with
$\left[{\cal E},\gamma_i,\gamma_o\right]$. While the
variation of ${\cal L}_j$ is quite sensitive on $\gamma_o$,
it is relatively less sensitive to the variation of
$\left[{\cal E},\gamma_i\right]$. Hence the thermal properties
of the outflow mainly contributes to the variation of ${\cal L}_j$ as is
expected.

Our approach to description of the low angular momentum flow is based on
several assumptions. 

First, in our work, viscous transport of the angular momentum is not explicitly
taken into account. Even
thirty years after the discovery of standard accretion disc theory
(Shakura \& Sunyaev 1973), exact modeling of viscous multi-transonic
BH accretion, including proper heating and cooling mechanisms is still
quite an arduous task. Nevertheless, extremely large radial velocity
close to the BH implies $\tau_{inf}<<\tau_{visc}$ ($\tau_{inf}$ and
$\tau_{visc}$ are the fall and the viscous time scales respectively). 
Large radial velocities even at larger distances are due to the fact 
that the angular momentum content of the accreting fluid 
is relatively low (see, e.g., Beloborodov \& Illarionov 1991, 
Igumenshchev \& Beloborodov 1997, Proga \& Begelman 2003). Hence,
our assumption of inviscid flow is not unjustified.

On the other hand, the introduction of
viscosity would further reduce the radial angular momentum of the accreting matter.
As we have 
seen (Fig. 3), lower values of $\lambda$ produces the smaller values of 
the outflow launching zone ($r_{\rm sh}$) and smaller amount of 
${\dot M}_{\rm out}$, which means (see Fig. 4 and
related discussions) that viscous 
transonic flow would produce flares from Sgr A$^*$ with  
shorter duration time scale ($\tau$), compared to the values of $\tau$ 
obtained using our inviscid flow treatment.
 
Second, we apply here the stationary solution to deduce the properties of the
time-dependent flow. However, explicit time-dependent considerations addressed so
far only specific issues like time-dependent and oscillatory behaviours of the
shock (see, e.g.
Okuda, Teresi, Toscano \& Molteni 2004, and references therein),
which, for example,  leads to the quasi-periodic
oscillation of black hole candidates (Das 2003, and references
therein). Several groups performed numerical hydrodynamical computations. For example,
numerical simulations by Coker \& Melia (1997) and Rockefeller et al. (2004) of winds 
from many stars lead 
to predictions of much higher
average angular momentum than adopted in our paper ($\lambda \sim 60$) but also to much higher
accretion rate than allowed by the Faraday rotation constraints (Baganoff et al. 2003a). 
The problem is that such computations
cannot resolve the flow deeply inside $R_{BHL}$, where the flare
formation occurs.  

Our picture of accretion flow is qualitatively similar to the one developed by Loeb 
(2004). However, he 
considered accretion from stars very close to the black hole, well within appropriate
Bondi-Hoyle-Lyttleton radius, with less efficient
winds and star velocities being smaller than the wind velocity only at a fraction of an orbit.
In this picture accretion from any of the considered stars is expected to occur once per orbit and
to last only for a few months.
IRS 13 E3 is expected to supply the mass continuously. Therefore, Loeb (2004) 
model would
require significant correlation of the long time-scale activity with the motion of
nearby stars while our picture would be in agreement with no such trends. Observational
search for such correlations will help to determine the dominant source of the accreting 
material.

\section*{Acknowledgments}  
 
This work was  
supported in part by grant 2P03D~003~22 (BCZ)
of the Polish State  
Committee for Scientific Research (KBN).
Work of TKD is supported by grants from KBN.
We gratefully acknowledge useful discussions with 
Mark Morris (UCLA) and Marek Sikora (CAMK).

\end{document}